\documentclass[twocolumn,showpacs,preprintnumbers,amsmath,amssymb]{revtex4}
\usepackage{graphicx}% Include figure files
\usepackage{dcolumn}% Align table columns on decimal point
\usepackage{bm}% bold math
\usepackage{color}
\begin{document}

%\preprint{APS/123-QED}

\title{Competing magnetic phases and field-induced dynamics in DyRuAsO}

\author{Michael A. McGuire}
\author{V. Ovidiu Garlea}
\author{Andrew F. May}
\author{Brian C. Sales}

\address{Oak Ridge National Laboratory, Oak Ridge, Tennessee 37831 USA}

\date{\today}

\begin{abstract}
Analysis of neutron diffraction, dc magnetization, ac magnetic susceptibility, heat capacity, and electrical resistivity for DyRuAsO in an applied magnetic field are presented at temperatures near and below those at which the structural distortion ($T_S$ = 25 K) and subsequent magnetic ordering ($T_N$ = 10.5 K) take place. Powder neutron diffraction is used to determine the antiferromagnetic order of Dy moments of magnitude 7.6(1) $\mu_B$ in the absence of a magnetic field, and demonstrate the reorientation of the moments into a ferromagnetic configuration upon application of a magnetic field.  Dy magnetism is identified as the driving force for the structural distortion. The magnetic structure of analogous TbRuAsO is also reported. Competition between the two magnetically ordered states in DyRuAsO is found to produce unusual physical properties in applied magnetic fields at low temperature. An additional phase transition near $T^*$ = 3 K is observed in heat capacity and other properties in fields $\gtrsim$ 3 T. Magnetic fields of this magnitude also induce spin-glass-like behavior including thermal and magnetic hysteresis, divergence of zero-field-cooled and field-cooled magnetization, frequency dependent anomalies in ac magnetic susceptibility, and slow relaxation of the magnetization. This is remarkable since DyRuAsO is a stoichiometric material with no disorder detected by neutron diffraction, and suggests analogies with spin-ice compounds and related materials with strong geometric frustration.

\end{abstract}

%\pacs{}
\maketitle

\section{Introduction}

The common crystallographic feature of layered iron superconductors is the Fe\textit{X} layer, composed of Fe bonded to \textit{X}, a pnictogen or chalcogen, in edge-sharing tetrahedral coordination. There are several related structural families of such compounds, which are differentiated from one another by what is found between the Fe\textit{X} layers \cite{Johrendt-Review, McGuire-review}. In many cases, isostructural compounds are known with other elements in place of Fe. Among 1111-type materials (ZrCuSiAs structure type) with compositions \textit{Ln}FeAsO (\textit{Ln} = trivalent lanthanide) \cite{Quebe-2000}, Fe can be fully replaced to form \textit{LnM}AsO where \textit{M} = Mn \cite{Nientiedt-LnMnAsO}, Ru \cite{Quebe-2000}, Co \cite{Quebe-2000}, Rh \cite{Muir-LaRhAsO}, Ir \cite{Muir-LnRhIrAsO}, Ni \cite{Watanabe-LaNiAsO}, Zn \cite{Nientiedt-LnZnAsO}, and Cd \cite{Charkin-LnCdAsO}.

The physical properties of \textit{LnM}AsO materials reflect a wide range of behaviors and associated ground states. These range from insulating to superconducting, and include many types of magnetism. \textit{Ln}MnAsO compounds are semiconductors displaying giant magnetoresistance and antiferromagnetic (AFM) ordering of Mn moments at temperatures often exceeding room temperature, and often accompanied by spin reorientation transitions at lower temperatures \cite{Marcinkova-NdMnAsO, Emery-LaNdMnAsO, Tsukamoto-CeMnAsO, Emery-LaNdMnAsO-2}. The well known Fe compounds \textit{Ln}FeAsO exhibit spin-density-wave-like AFM which is at least partly itinerant, and is strongly coupled to the lattice distortion which occurs near the magnetic ordering transition\cite{delaCruz-LaFeAsO, Johnston-review, Lumsden-Christianson-review, McGuire-review}.  Ferromagnetism (FM) associated with itinerant magnetic moments on Co is observed in \textit{Ln}CoAsO, and a transition to AFM order is observed at lower temperatures due to effects of localized 4\textit{f} moments on the magnetic lanthanides \cite{Yanagi-LaCoAsO, Ohta-LCoAsO, McGuire-NdCoAsO, Marcinkova-NdCoAsO}. Studies of the 4\textit{d} and 5\textit{d} transition metal analogues of the Co compounds (\textit{Ln}RhAsO and \textit{Ln}IrAsO) report only rare-earth magnetic ordering and only in the case of \textit{Ln} = Ce \cite{Muir-LaRhAsO, Muir-LnRhIrAsO}. Some \textit{Ln}NiAsO compounds are superconducting at low temperatures ($Ln$ = La\cite{Watanabe-LaNiAsO, Li-LaNiAsO}, Pr \cite{Matsuishi-LnNiAsO}), while others are not. CeNiAsO shows two magnetic ordering transitions associated with Ce moments, and is described as a dense Kondo lattice metal \cite{Luo-CeNiAsO}. With a filled 3$d$ shell, $Ln$ZnAsO compounds are semiconductors with band gaps near 2 eV, and form as transparent crystals with colors varying from yellow-orange to red \cite{Lincke-LnZnAsO}. Clearly this structure type provides fertile grounds for interesting physics accessible by simple chemical substitutions.

Among the many studied substitutions, Ru is unique in that it is isoelectronic with Fe. Partial replacement of Fe with Ru in the related 122 materials SrFe$_2$As$_2$ and BaFe$_2$As$_2$ produces superconductivity with transition temperatures near 20 K \cite{Schnelle-SrFeRuAs, Sharma-BaFeRuAs}, while partial substitution of Ru into 1111-type materials only suppresses the magnetism without the appearance of superconductivity \cite{McGuire-PrFeRuAsO, Pallecchi-LaFeRuAsO}. \textit{Ln}RuAsO compounds are metals, and show magnetic ordering at low temperatures when magnetic lanthanides are included \cite{Chen-LnRuAsO, McGuire-LnRuAsO, McGuire-TbDyRuAsO}.

Our previous study of Ru containing 1111 materials uncovered particularly interesting behaviors in DyRuAsO \cite{McGuire-TbDyRuAsO}. This material undergoes a structural phase transition from tetragonal to orthorhombic near $T_S$ = 25 K, but adopts a different low temperature structure (Fig. \ref{fig:MagStr}c) than that observed in $Ln$FeAsO. The distortion which occurs in DyRuAsO involves a stretching of the unit cell along the $a$-axis, maintaining a single Ru$-$Ru distance within the Ru net, but Ru$-$Ru$-$Ru angles which deviate from 90$^{\circ}$. This is unlike the distortion which occurs in the parent phases of the layered iron superconductors, which shears the unit cell and results in a rectangular net of Fe atoms \cite{delaCruz-LaFeAsO}. In addition, anomalies in heat capacity and magnetization of DyRuAsO indicate magnetic ordering below $T_N$ = 10.5 K.

The temperature and magnetic field dependence of the physical properties indicated complex physics related to magnetism is at play in DyRuAsO, and indications of strong magnetoelastic coupling were observed. A metamagnetic transition was observed below $T_N$, and the heat capacity anomaly at the structural transition responded strongly to a magnetic field. The present work aims to improve our understanding of the underlying physics related to these phenomena by identifying the magnetic structures, determining the possible role of Ru magnetism, and further examining the effects of the competition between ordered ground states on the thermal, transport, and magnetic properties of this material.

Here we report results of neutron powder diffraction experiments which reveal the low temperature magnetic orderings, and the nature of the field induced transition, which involves competing AFM and FM phases. Effects of the competition between these phases include thermal and magnetic hysteresis in the magnetization and electrical resistivity, divergence between zero field cooled and field cooled magnetization data, frequency dependence of the dynamical (ac) susceptibility, and time dependent properties over a range of magnetic fields and temperatures. These behaviors highlight the complexity of the magnetic interaction in DyRuAsO, and are reminiscent of spin-glass physics under some conditions. Dy magnetism is identified as the driving force for the structural phase transition, with little or no influence from Ru. In addition, detailed heat capacity, electrical resistivity, and magnetization measurements provide evidence of a third phase transition which appears to be related to the competing magnetic ground states. The transition occurs near $T^*$ = 3 K and is strongest at magnetic fields of 3$-$4 T, diminished at higher fields, and absent at fields of 0$-$2 T.

\section{Experimental Details}

Polycrystalline samples were synthesized from Dy, Dy$_2$O$_3$, and RuAs as described in Ref. \citenum{McGuire-TbDyRuAsO}. Rietveld refinement of powder diffraction patterns (neutron and x-ray) of the samples used in this study show them to be $\gtrsim$95\% pure, with Dy$_2$O$_3$ as the main impurity. Neutron diffraction experiments were conducted at the High Flux Isotope Reactor at Oak Ridge National Laboratory using the Neutron Powder Diffractometer (beamline HB-2A). Data were collected at multiple temperatures and magnetic fields and using two neutron wavelengths, 1.538 {\AA} and 2.41 {\AA}. A collection of rectangular bars ($4\times4\times7$ mm$^3$) were cut from a sintered polycrystalline pellet for the neutron diffraction measurement to prevent rotation of the powder grains in the applied magnetic field. A vanadium can with inner diameter of 6 mm was used to contain the sample, which was loaded into a 5 T vertical-field cryomagnet. The field was directed perpendicular to the scattering plane. Dy has a high thermal neutron absorption cross section; however, this did not preclude the collection of data of sufficient quality for Rietveld analysis, which was performed using FullProf \cite{Fullprof}. Similar neutron powder diffraction measurements, with no applied magnetic field, were also performed on TbRuAsO, also prepared as described in Ref. \citenum{McGuire-TbDyRuAsO}.

Measurements of the temperature and magnetic field dependence of the electrical resistivity, ac and dc magnetization, and heat capacity were performed using a Quantum Design Physical Property Measurement System. Electrical contacts were made using platinum wires and conducting silver paste.

\section{Results and Discussion}

\subsection{Magnetic structure}

\begin{figure}
\begin{center}
\includegraphics[width=3.25in]{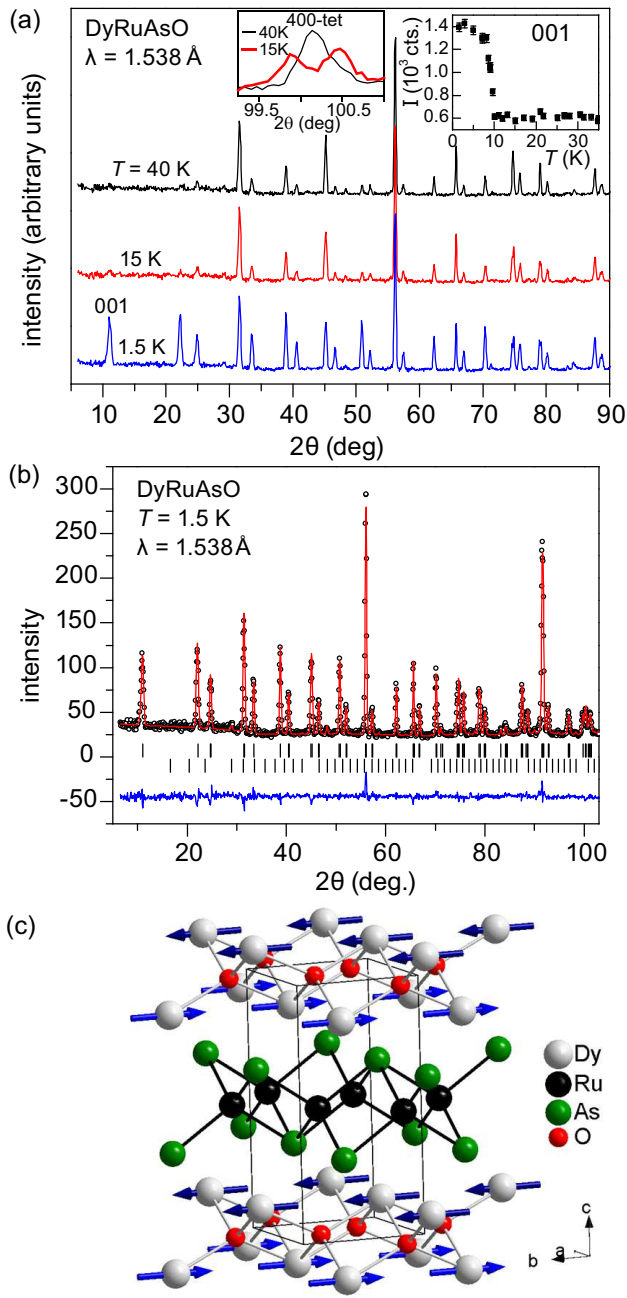}
\caption{\label{fig:MagStr}
(Color online) (a) Neutron powder diffraction patterns from DyRuAsO collected in the tetragonal state at 40 K, the orthorhombic state at 15 K, and the magnetically ordered state at 1.5 K. The insets in (a) show the splitting of the tetragonal 400 reflection resulting from the structural transition and the intensity of the 001 reflection (labeled in the main panel) as a function of temperature, illustrating the onset of magnetic order below about 10 K. (b) Rietveld fits of the nuclear and magnetic structures of DyRuAsO at 1.5 K. The lower ticks locate reflections from the Dy$_2$O$_3$ impurity. The AFM arrangement of the Dy moments determined from the diffraction analysis is shown in (c).
}
\end{center}
\end{figure}

Figure \ref{fig:MagStr} shows neutron powder diffraction (NPD) data collected in the paramagnetic, tetragonal state at 40 K, in the paramagnetic, orthorhombic state at 15 K, and in the magnetically ordered, orthorhombic state at 1.5 K. Rietveld analysis (not shown) of the patterns at 40 K and 15 K are consistent with the tetragonal and orthorhombic structures \cite{McGuire-TbDyRuAsO}, respectively, with no indication of mixed or partial occupancy of any of the atomic sites. No indications of long range magnetic order accompanying the structural transition is seen. Strong magnetic reflections are observed at 1.5 K. All of the magnetic scattering occurs at the positions of nuclear Bragg reflections, indicating the magnetic and nuclear unit cells are identical [propagation vector \textbf{k} = (0 0 0)]. The temperature dependence of the intensity of the 001 reflection is shown in the inset of Figure \ref{fig:MagStr}. The onset of magnetic order occurs near 10 K, consistent with magnetization measurements which identify $T_N$ = 10.5 K in Ref. \citenum{McGuire-TbDyRuAsO} and below, with saturation of the ordered moment occurring near 5 K.

Representational analysis was used to determine the symmetry-allowed magnetic structures that can result from a second-order magnetic phase transition, given the crystal symmetry before the transition ($Pmmn$) and the propagation vector of the magnetic ordering, \textbf{k}=( 0 0 0) . These calculations were carried out using the program SARA{\it h}-Representational Analysis.\cite{Wills-Sarah} The decomposition of the magnetic representation (i.e. irreducible representations (IRs))  for the Dy site $( 0.25,~ 0.25,~ z)$ is $\Gamma_{Mag}=1\Gamma_{2}+1\Gamma_{3}+1\Gamma_{4}+1\Gamma_{5}+1\Gamma_{6}+1\Gamma_{7}$. The labeling of the propagation vector and the IRs follows the scheme used by Kovalev\cite{Kovalev}. Each representation contains only one basis vector meaning that the magnetic moments are confined to one direction, while the two Dy atoms of the primitive cell can carry parallel or antiparallel moments. Strong magnetic contributions to the 00$\ell$ reflections indicate moments with large components in the \textit{ab}-plane. Rietveld refinement results of the diffraction data collected below $T_N$ using wavelength 1.538 {\AA} are shown in Figure \ref{fig:MagStr}(b). The orthorhombic distortion (\textit{a} $>$ \textit{b}) allows the distinction of the directions in the \textit{ab}-plane, and the best fits were obtained with the Dy moments along the \textit{b}-axis, corresponding to the representation $\Gamma_4$ (or Shubnikov magnetic space group $Pm'mn$). In this model, the moments on the two Dy atoms in the primitive cell are aligned AFM. This produces FM layers of Dy in the \textit{ab}-plane, with AFM alignment between neighboring Dy layers.The resulting magnetic structure is shown in Figure \ref{fig:MagStr}(c). Because of the compression of the ODy$_4$ tetrahedral units along the \textit{c}-axis (Fig. \ref{fig:MagStr}c), the shortest Dy$-$Dy distances (3.5 {\AA}) are those between neighboring layers within the DyO slabs. The shortest distance between Dy atoms within a single layer are considerably longer (4.0 {\AA}).

Neutron powder diffraction analysis of isostructural TbRuAsO was also performed, and the same magnetic structure as DyRuAsO was determined for the Tb moments. In this case, however, no distinction between the \textit{a} and \textit{b} directions can be made, since TbRuAsO remains tetragonal within experimental resolution to at least 1.5 K. Structural information and magnetic moments determined from the refinements of both compounds at the lowest temperatures investigated are collected in Table \ref{tab:refinements}. No conclusive evidence for ordered magnetic moments on Ru is seen in the data. Small, non-zero values ($\lesssim$ 0.5 $\mu_B$) are obtained when Ru moments are included in the refinements at the lowest temperatures, but the quality of the fit is not improved by their addition. The refined values of the rare-earth moment at 1.5 K are 7.6(1) $\mu_B$ for Dy in DyRuAsO and 5.76(8) $\mu_B$ for Tb in TbRuAsO.

The refined ordered moments are reduced from their free ion values of \textit{gJ}, which are 9 $\mu_B$ for Tb and 10 $\mu_B$ for Dy, as commonly found in related Fe-based materials \cite{CeFeAsO-moment, PrFeAsO-moment, NdFeAsO-moment, NdFeAsO-CEF}. This is attributed to crystalline electric field effects, the details of which are not known at this time. In these materials the rare earths are in somewhat irregular coordination environments. The nearest neighbors of the Dy and Tb sites form distorted square-antiprisms, with the squares formed by As on one side and O on the other (distorted squares in the case of orthorhombic DyRuAsO). At 1.5 K the site symmetry is \textit{mm2} for Dy (Wyckoff position 2\textit{a}) and 4\textit{mm} for Tb (Wyckoff position 2\textit{c}). It is likely that the temperature and magnetic field dependence of the relative positions of the crystal field levels plays an important role in the unusual magnetic properties described below.

\begin{table}
\begin{center}
\caption{\label{tab:refinements} Results and agreement factors from Rietveld refinement of neutron ($\lambda$ = 1.538 {\AA}) powder diffraction data for DyRuAsO and TbRuAsO at 1.5 K with no applied magnetic field. Dy/Tb and As occupy sites at (1/4, 1/4, \textit{z}), while Ru and O occupy sites at (3/4, 1/4, \textit{z}).}
\begin{tabular}{lcc}
\hline
	&	DyRuAsO	&	TbRuAsO	\\	\hline
space group	&	\textit{Pmmn}	&	\textit{P}4/\textit{nmm}	\\	
a  ({\AA})	&	4.0222(1)	&	4.0215(1)	\\	
b  ({\AA})	&	4.0070(1)	&	= a	\\	
c  ({\AA})	&	8.0092(3)	&	8.0558(3)	\\	
$z_{Dy/Tb}$	&	0.1311(5)	&	0.1332(8)	\\	
$z_{Ru}$	&	0.500(2)	&	1/2	\\	
$z_{As}$	&	0.665(1)	&	0.664(1)	\\	
$z_{O}$	&	0.013(2)	&	0	\\	
$m_{Dy/Tb}$ ($\mu_B$)	&	7.6(1)	&	5.76(8)	\\	
$\chi^2$	&	1.03	&	3.56	\\	
$R_{mag}$	&	8.29	&	4.77	\\	\hline
\end{tabular}
\end{center}
\end{table}

\begin{figure}
\begin{center}
\includegraphics[width=3.25in]{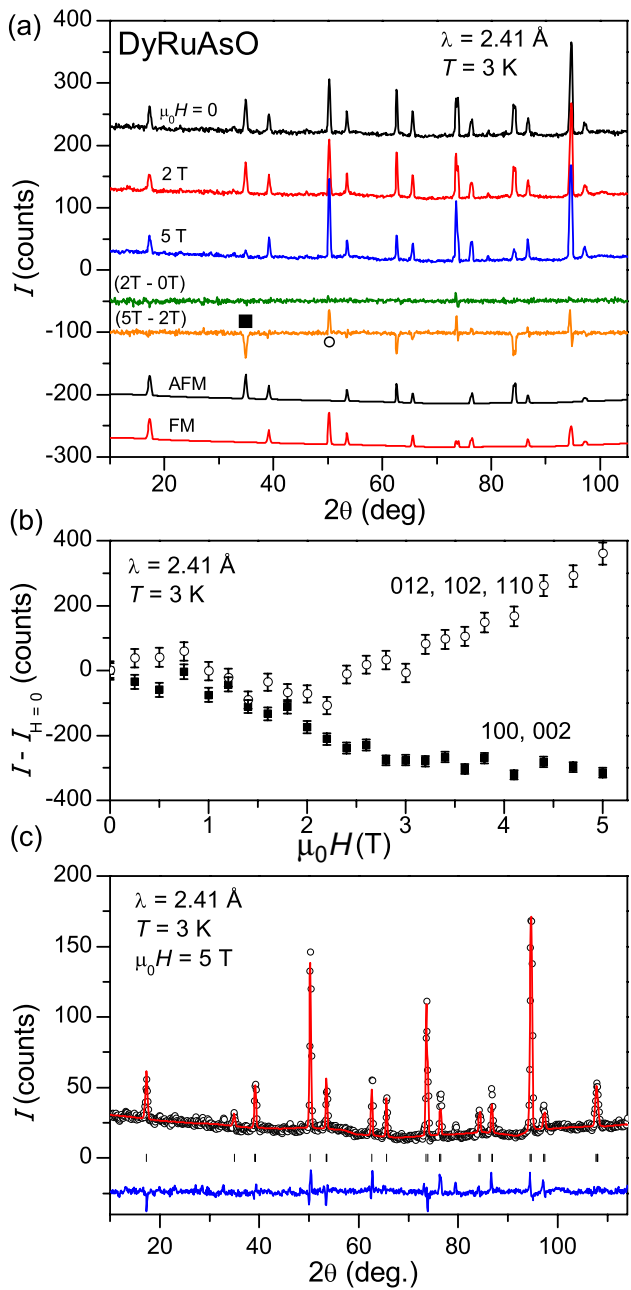}
\caption{\label{fig:NPDvsH}
(Color online) (a) Neutron powder diffraction patterns collected at 3 K with applied magnetic field of $\mu_0H$ = 0, 2, and 5 T. The difference between the data collected at 2 T and zero field, and between the data collected at 5 T and 2 T are also shown. The lowest two patterns are simulations including only magnetic scattering for AFM and FM moments on Dy. Patterns are offset vertically for clarity. (b) The field dependence of the relative scattered intensity at the peaks marked by the square and circle in (a), showing a divergence for fields above about 2 T. The data are labeled by Miller indices of the overlapping reflections which contribute magnetic scattering intensity to the measured peaks. (c) Rietveld refinement results for $\mu_0H$ = 5 T and T = 3 K using a predominately FM model (see text for details) with all Dy moments along the b-axis.
}
\end{center}
\end{figure}

Since evidence for a metamagnetic transition and a strong magnetic field effect on the heat capacity has been observed in DyRuAsO \cite{McGuire-TbDyRuAsO}, NPD data were also collected in applied magnetic fields. Results of these experiments are shown in Figure \ref{fig:NPDvsH}. As seen in the difference curves in the middle of Fig. \ref{fig:NPDvsH}a, application of a 2 T magnetic field at \textit{T} = 3 K has little effect on the diffracted intensities; however, significant changes are observed as the field is increased to 5 T. It is important to note that many reflections show little or no response to the magnetic field. This shows that the texture of the pelletized sample is not affected. The magnetic field dependence of the intensity of two diffraction peaks, measured upon decreasing the magnetic field, is shown in Fig. \ref{fig:NPDvsH}b. These peaks are labeled by the Miller indices of the reflections which contribute magnetic scattering intensity to them, and are identified in Fig. \ref{fig:NPDvsH}a by the data markers used in Fig. \ref{fig:NPDvsH}b. There is a clear change in the field dependence which onsets near 2 T, which is identified as a transition from AFM to FM order. At the bottom of Fig. \ref{fig:NPDvsH}a, simulated diffraction patterns including only the magnetic contribution are shown for the AFM structure determined at zero applied field (Fig. \ref{fig:MagStr}c) and for the FM structure obtained by aligning all the Dy moments along the b-axis (corresponding to IR $\Gamma_5$ and Shubnikov group $Pm'mn'$). Comparing these simulations with the difference curve between the 5 T and 2 T data reveals that the peaks which are strongly suppressed at high fields are associated only with the AFM structure and those which are enhanced at high field are associated only with the FM structure.

Results from Rietveld refinement of data collected at $\mu_0H$= 5 T and \textit{T} = 3 K are shown in Fig. \ref{fig:NPDvsH}c. The majority of the magnetic scattering is accounted for using the FM model with Dy moments of 7.3(3) $\mu_B$ along the b-axis; however, the data indicates the presence of a small AFM component as well. The fit shown includes both FM and AFM contributions, and the fraction of the AFM phase is estimated to be about 13\% at 5 T. The refinement is relatively insensitive to the direction of the moment within \textit{ab}-plane, and similar results are obtained when the FM moment is constrained along the \textit{b}-axis, or allowed to have components along both \textit{a} and \textit{b}. No indication of a \textit{c}-component is observed. The transition between the AFM and FM structures involves a change in the relative orientation of moments on nearest neighbor Dy atoms. In addition, the data is not consistent with a fully polarized powder, in which every grain, regardless of crystallographic orientation, would have a moment directed perpendicular to the scattering plane. These results are in agreement with the magnetic properties discussed below, in which a preference for moments in the \textit{ab}-plane is inferred, and a field of 5 T is seen to be insufficient to fully polarize the polycrystalline material.

Similar fits (not shown) were performed for data collected at fields of 2 and 5 T and temperatures of 15 K (below the structural transition) and 40 K (above the structural transition). Of these, only the pattern from 15 K at 5 T indicated the presence of magnetic scattering, which was well modeled with FM ordering of 5.5(3) $\mu_B$ moments on Dy aligned along the b-axis. This temperature is above $T_N$, and no indication of an AFM component was observed at any field at this temperature. This shows that FM order emerges out of the orthorhombic, paramagnetic state ($T_S > T > T_N$) when a large magnetic field is applied.

It is expected that the competing FM and AFM states may strongly affect the physical properties of DyRuAsO, and consideration of this competition is required in understanding the behavior of magnetic, transport, and thermal properties presented below.

\subsection{dc magnetization}

Figure \ref{fig:dcmag}a shows the results of dc magnetization measurements as a function of applied field for a wide range of temperatures. Similar results restricted to lower fields and fewer temperatures were previously reported \cite{McGuire-TbDyRuAsO}. At temperatures below $T_N$, a rapid increase in the magnetic moment ($m$) is observed as the field is increased beyond 2$-$3 T. This is consistent with the analysis of the neutron diffraction data presented above. At 2 K, the magnetic moment approaches a saturation value of 6.8 $\mu_B$ per formula unit at 12 T similar to but less than the ordered moment on Dy of 7.6(1) $\mu_B$ determined by neutron diffraction in zero applied field. An approach to saturation near the same value can be inferred from the data above $T_N$ in the paramagnetic state as well, as expected for large moments in high fields at relatively low temperatures.

It is interesting to compare the measured magnetization with the results of the neutron diffraction measurements. At 3 K, in a field of 5 T, the refined value of the FM moment on Dy is 7.3(3) $\mu_B$, and the data suggest the moments are constrained to lie in the \textit{ab}-plane. The measured magnetic moment at this temperature and field is 5.1 $\mu_B$ per Dy. This is about 2/3 of the refined moment. Such a suppression of the measured moment relative to the ordered moment is expected due to the magnetic anisotropy; some crystallites in the magnetization sample will have their \textit{c}-axes along the field direction, and thus not contribute to the measured magnetization.

\begin{figure}
\begin{center}
\includegraphics[width=3.25in]{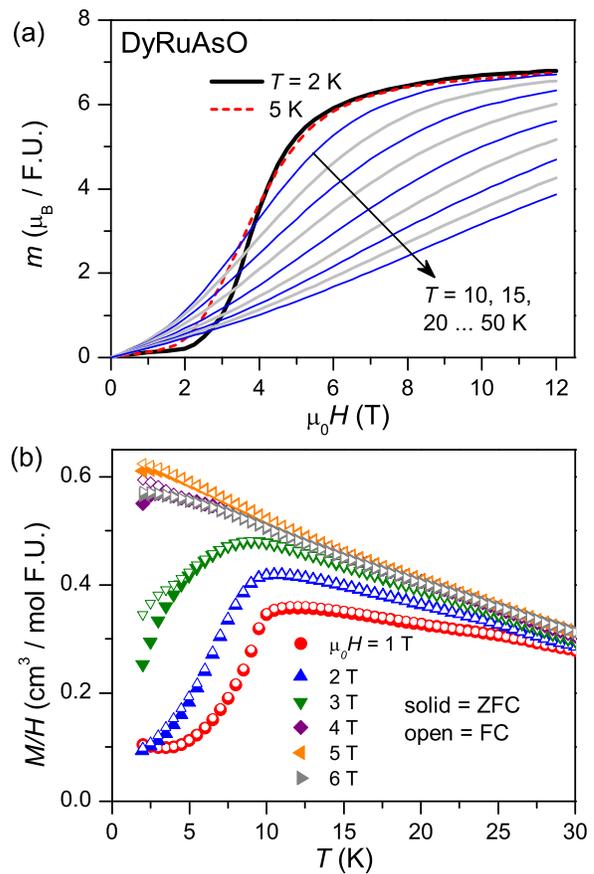}
\caption{\label{fig:dcmag}
(Color online) (a) The field dependence of the magnetic moment per formula unit determined from dc magnetization measurements. (b) Low temperature behavior of M/H measured on warming after zero field cooling (ZFC) and field cooling (FC), showing a divergence which is strongest at intermediate fields.
}
\end{center}
\end{figure}

Results of dc magnetization vs. temperature measurements are summarized in Fig. \ref{fig:dcmag}b, which shows the temperature dependence of \textit{M/H} at low temperatures in applied fields ranging from 1 to 6 T. Data were collected using both zero field cooled (ZFC) and field cooled (FC) procedures. At temperatures above about 50 K, similar behavior is observed in all of the applied fields.  The large decrease in \textit{M/H} upon cooling through $T_N$ for $\mu_0H< 3 T$ in Figure \ref{fig:dcmag}b is noteworthy. At the lowest applied fields, the moment decreases by approximately two-thirds relative to the value observed just above $T_N$. In a typical AFM, the powder-averaged moment decreases by only one third below $T_N$ \cite{Smart}. This is indicative of anisotropic susceptibility in the paramagnetic state, with a larger than average value in the direction along which the moments order below $T_N$, the \textit{b}-axis in this case. Since the orthorhombic distortion is small, it may be expected that $\chi_a \approx \chi_b$, which then would imply that $\chi_c$ is small compared to the in-plane susceptibility. This suggests that the Dy moments prefer to lie in the \textit{ab}-plane above $T_N$, as well as in the ordered state. Similar easy-plane anisotropy has been observed in the paramagnetic state of CeFeAsO \cite{Jesche-CeFeAsO}, which has ordered Ce moments lying nearly in the ab-plane at low temperatures \cite{Zhao-Ce}. A divergence between the ZFC and FC data is observed in Figure \ref{fig:dcmag}b near 5 K for magnetic fields larger than 2 T, associated with the emerging FM.

As previously noted \cite{McGuire-TbDyRuAsO}, an anomaly occurs in \textit{M/H} at $T_S$, most easily seen at low fields. No long ranged magnetic order is seen in the diffraction data, so this could signal a change in the crystal field levels of the Dy ion accompanying the structural distortion. This could explain why the effective moment determined by the Curie-Weiss model in Ref. \citenum{McGuire-TbDyRuAsO} agree well with the expectations for the free Dy ion, while the ordered moment at low temperatures does not. The evolution of the crystal field levels with temperature and magnetic field below this structural phase transition is expected to be complex, but may prove important in understanding this material.

\subsection{Heat capacity}

\begin{figure}
\begin{center}
\includegraphics[width=3.25in]{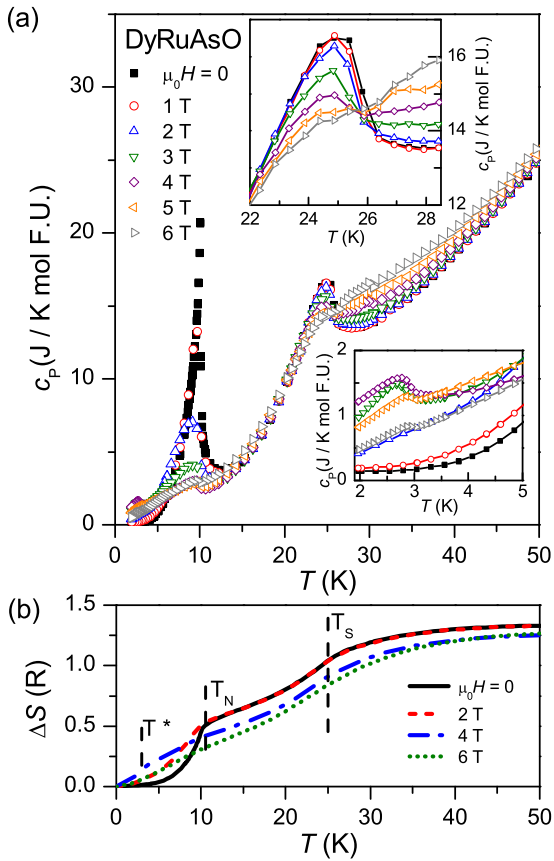}
\caption{\label{fig:hc}
(Color online) (a) Heat capacity of DyRuAsO at the indicated magnetic fields. The insets shows the behaviors near $T_S = 25 K$ and $T* = 3 K$. (b) The entropy change determined by integration of $c_P/T$, after subtracting a estimated background contribution determined by scaling data from LaRuAsO to match the data measured at zero field and 50 K.
}
\end{center}
\end{figure}

Application of a 6 T magnetic field has been shown to affect strongly the heat capacity anomalies at $T_S$ and $T_N$ \cite{McGuire-TbDyRuAsO}. Heat capacity data collected at $\mu_0H$= 0$-$6 T in 1 T increments are shown in Fig. \ref{fig:hc}a. As the magnetic field is increased, the peak at $T_N$ is gradually suppressed and broadened toward lower temperatures, with only a small anomaly remaining at 6 T. The insets show the behaviors near $T_S$ (upper inset) and at low temperatures (lower inset). For fields up to 2 T, the heat capacity anomaly at $T_S$ is nearly unchanged. When the field is increased to 3 T and above, the peak is suppressed and skewed toward higher temperatures. The suppression and skewing increase with field for $\mu_0H\geq 3 T$. This field effect on the anomaly at $T_S$ indicates a magnetic component to the phase transition occurring at this temperature, or at minimum supports strong magnetoelastic coupling. The field dependence is similar to that expected for FM ordering, although no long range ordering above $T_N$ is discernable in the neutron diffraction data discussed above. To gain some insight into the origin of the structural distortion, samples of the transition-metal-free analog DyZnAsO were synthesized and preliminary heat capacity and diffraction measurements were performed (see Supplemental Material). The tetragonal to orthorhombic distortion indeed occurs in DyZnAsO at a temperature near 30 K. This eliminates Ru magnetism or orbital ordering as a source for the structural distortion, and implicates Dy magnetism.

At the lowest temperatures (Fig. \ref{fig:hc}a, lower inset), the effect of increasing the field from 0 to 2 T is an overall increase in magnitude, caused at least in part by the broadening of the anomaly at $T_N$. However, a qualitative change in the low temperature heat capacity occurs as $\mu_0H$ is increased to 3 T, as also noted near $T_S$. At this field an additional peak appears below $T^*$ = 3 K, indicating an additional phase transition in this material. The strong field response suggests that this transition is magnetic in nature, and appears to be related to several unusual behaviors of the magnetic and transport properties which will be discussed below.

The entropy change ($\Delta S$) up to 50 K, estimated by integration of $c_P/T$ after subtraction of a background $c_P$, are shown in Fig. \ref{fig:hc}b. The background data were estimated by scaling the heat capacity of LaRuAsO \cite{McGuire-LnRuAsO} to match the measured heat capacity of DyRuAsO at zero field and 50 K. Isostructural LaRuAsO is not known to undergo any magnetic or crystallographic phase transitions in this temperature range. For purposes of integration, the background-subtracted $c_P/T$ data were linearly interpolated from the lowest measurement temperature to the origin. In all of the studied magnetic fields, the total entropy change up to 50 K is similar. Although increasing the field from zero to 2 T significantly suppresses the sharpness of the peak at $T_N$, the total entropy associated with it is not changed and is about 80\% of $R$ln(2). Increasing the field beyond 2 T results in a suppression of the entropy obtained upon integration up to $T_N$. The change in $\Delta S$ between $T_N$ and 50 K is similar in all the fields studied (0.8$-$0.9$R$). This suggests the total entropy associated with the structural transition at $T_S$ not strongly dependent on the applied field, though the shape of the anomaly is significantly changed.

The phase transitions observed in DyRuAsO appear to be second order in nature. This is indicated by the shapes of the heat capacity anomalies in Fig. \ref{fig:hc}a, the lack of any anomalous behavior in the raw heat capacity data \cite{Lashley}, the temperature dependence of the magnetic order parameter (Fig. \ref{fig:MagStr}a) and the absence of thermal hysteresis in the physical properties measured at zero field.

\subsection{Magnetoresistance}

\begin{figure}
\begin{center}
\includegraphics[width=3.25in]{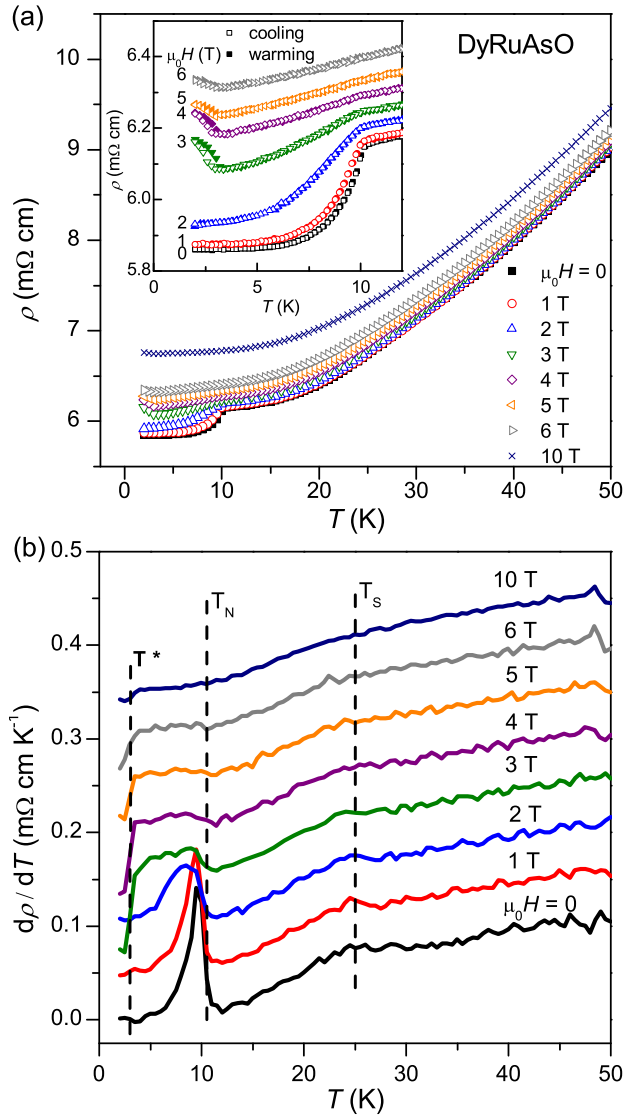}
\caption{\label{fig:res}
(Color online) (a)Resistivity ($\rho$) below 50 K in the main panel, with the behavior near and below $T_N$ shown in the inset. Data were collected on cooling in the indicated applied field (open markers), followed \emph{immediately} by warming in the same field (solid markers). (b) Temperature derivative of $\rho$ with phase transition temperatures marked on the plot. The curves in (b) are offset vertically for clarity.
}
\end{center}
\end{figure}

Magnetic field effects on the temperature dependence of the electrical resistivity ($\rho$) of DyRuAsO are depicted in Fig. \ref{fig:res}. An abrupt decrease in $\rho$ is observed upon cooling through $T_N$ for $\mu_0H$$\lesssim$ 2T. This feature is diminished significantly at higher fields. The effect of the structural transition is not directly apparent from the observed temperature dependence; however, a slope change in $d\rho/dT$ is seen at $T_S$ (Fig. \ref{fig:res}b). Though it is subtle, this anomaly persists up to 10 T, suggesting the structural transition occurs at all fields investigated here.

Effects of the field induced transition at $T^*$ are also observed in the resistivity data. This is manifested as an abrupt upturn in $\rho$ occurring below 3 K for $\mu_0H\geq$ 3 T (Fig. \ref{fig:res}a). This is clearly observed in the derivative (Fig. \ref{fig:res}b). The upturn in $\rho$ is strongest for $\mu_0H$= 3 and 4 T, the fields at which the heat capacity anomaly at $T^*$ is also strongest. In the inset of Fig. \ref{fig:res}a, $\rho$ data are shown for cooling in the applied field followed immediately by warming in the applied field. In addition to the upturn upon cooling already noted, an apparent thermal hysteresis is observed below $T^*$ for $\mu_0H\geq$ 3 T, and is observed most clearly at 3 T. In the following discussion, this will be shown to be related to slow relaxation of the magnetic state of the material upon moving through that particular region of the $H-T$ phase diagram.

\subsection{Time and frequency dependent phenomena}

\begin{figure}
\begin{center}
\includegraphics[width=3.0in]{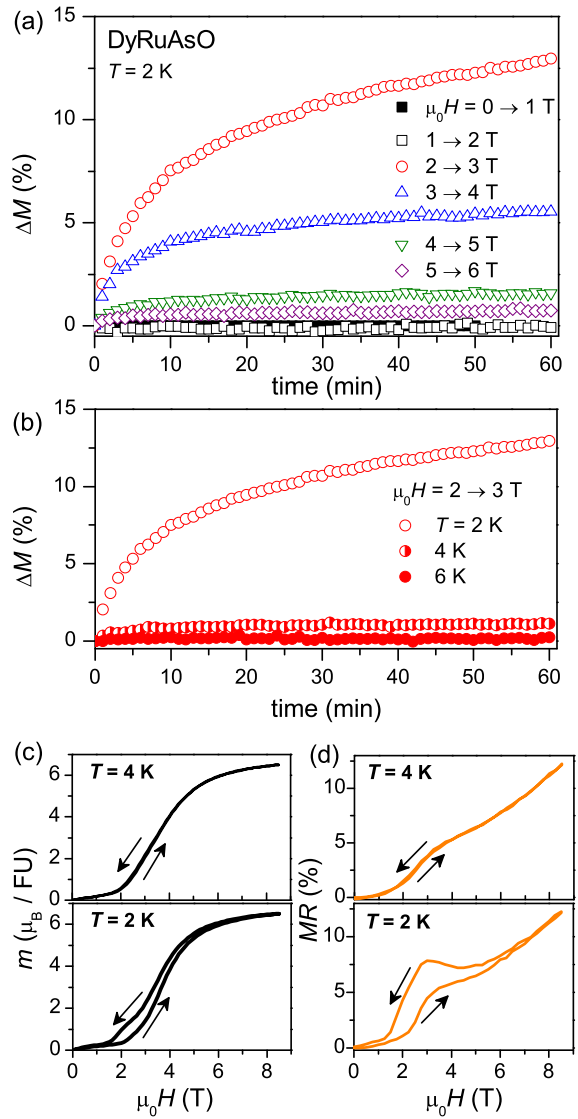}
\caption{\label{fig:relaxation}
(Color online) (a) Time dependence of the magnetic moment of DyRuAsO after increasing the field as indicated in the legend at a temperature of 2 K. (b) Time dependence of the magnetic moment after increasing the field from 2 to 3 T at T = 2, 4 and 6 K. (c) Field dependence of the magnetic moment measured upon increasing then decreasing the field at T = 2 and 4 K. (d) Field dependence of the magnetoresistance relative to the zero field resistivity values measured upon increasing then decreasing the field at T = 2 and 4 K.
}
\end{center}
\end{figure}

Competition between the AFM state stable at low fields and the FM state stable at high fields leads to several unusual properties occurring near the associated critical temperatures and magnetic fields. The ZFC-FC divergence in dc magnetization below about 5 K (Fig. \ref{fig:dcmag}b) and the apparent thermal hysteresis in the electrical resistivity (Fig. \ref{fig:res}a) were noted above when the applied field was $\geq$ 3 T. The dynamics in DyRuAsO were further examined by measurement of the time dependence of the magnetization, with results shown in Fig. \ref{fig:relaxation}. The sample was cooled to 2 K in zero field, and then the field was increased in 1 T increments up to 6 T. At each field, the magnetic moment ($M$) was measured every minute for 60 minutes. The time required to ramp and stabilize the field was approximately 3 minutes. The data are shown in Fig. \ref{fig:relaxation}a, plotted as a percentage difference relative to the value measured just after the magnetic field stabilized ($t =0$). When the field is increased from 0$\rightarrow$1 T and 1$\rightarrow$2 T, no relaxation of the magnetization is observed; the measured moment is independent of time. However, when the field is increased further, a time dependent moment is observed. The time dependence is strongest after increasing the field from 2$\rightarrow$3 T and 3$\rightarrow$4 T. A small change with time occurs for higher fields as well.

The time evolution of the moment was also examined at higher temperatures. The results after increasing the field from 2$\rightarrow$3T at 2, 4, and 6 K are shown in Fig. \ref{fig:relaxation}b. The relaxation seen at 2 K is strongly suppressed, but still observable, at 4 K and absent at 6 K. Similar behavior is seen in the magnetoresistance (not shown). This is likely related to the ZFC-FC divergence in the magnetization (Fig. \ref{fig:dcmag}b) and the divergence of the resistivity measured upon cooling and then warming (Fig. \ref{fig:res}a) when the field is 3 T or higher.

The dynamics resulting from the competition between the AFM and FM states is also demonstrated in the magnetic moment and magnetoresistance ($MR$) measured at fixed temperature upon increasing and then decreasing the field. Figure \ref{fig:relaxation}c shows the field dependence of the moment (field sweep rate of 15 Oe$\cdot$s$^{-1}$), which displays a divergence which is strongest between about 2 and 4 T at 2 K, but no detectable divergence at 4 K. Similar results are seen for the magnetoresistance (field sweep rate of 25 Oe$\cdot$s$^{-1}$) in Fig. \ref{fig:relaxation}d. The local maximum in $MR$ upon decreasing the field at 2 K indicates a significant enhancement in charge carrier scattering under these conditions.

\begin{figure}
\begin{center}
\includegraphics[width=3.25in]{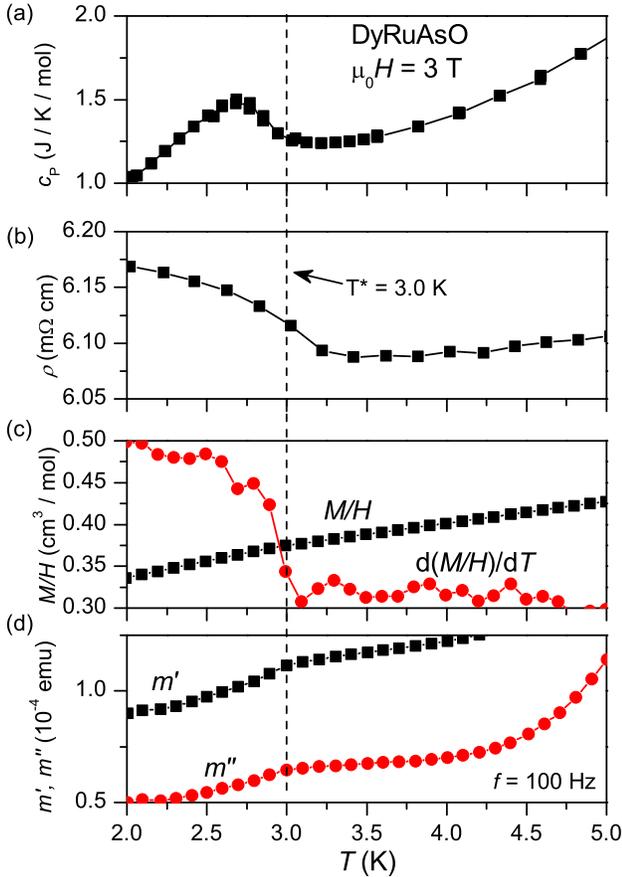}
\caption{\label{fig:3T}
(Color online) (a) Heat capacity, (b) electrical resistivity, (c) dc ``susceptibility'' ($M/H$), and (d) ac magnetization of DyRuAsO near $T^*$ in an applied magnetic field of 3 T. In (c) the temperature derivative of \textit{M/H} is also shown, in arbitrary units.
}
\end{center}
\end{figure}

The observation of slow dynamics at intermediate magnetic fields for temperatures below $T^*$ but not above (Fig. \ref{fig:relaxation}), suggests that the dissipation is related to the changes in other physical properties near in this temperature and field range. For comparison, thermal, transport, and magnetic properties measured at 3 T near $T^*$ are re-plotted together in Fig. \ref{fig:3T}. These anomalies suggest a phase transition occurs near this temperature for magnetic fields exceeding about 2 T, the field above which the FM phase fraction appears to increase most rapidly at low temperatures. The heat capacity (Figs. \ref{fig:3T}a and \ref{fig:hc}) shows a small but relatively sharp anomaly at $T^*$ for $\mu_0H$= 3$-$5 T. It is significantly suppressed at higher fields. The electrical resistivity (Figs. \ref{fig:3T}b and \ref{fig:res}) increases sharply upon cooling through this transitions in fields greater than 3 T. This behavior is still clearly observed at 6 T, but is suppressed at 10 T. In the dc magnetization (Fig. \ref{fig:dcmag}), the strong signal from Dy moments overwhelm subtle features at low temperature, but a marked FC-ZFC divergence onsets just above $T^*$ in this same range of magnetic fields. In addition, a subtle downturn is observed in FC data at $T^*$ for $\mu_0H$= 3 T (Fig. \ref{fig:3T}c). Anomalies near $T^*$ are observed in both components of the ac susceptibility at $\mu_0H$= 3 T (Fig. \ref{fig:3T}d), which is presented in more detail below.

The shape and relative sharpness of the heat capacity anomaly at $T^*$ and the observation of anomalies at this temperature in other physical properties are indicative of a thermodynamic phase transition, and not, for example, a Schottky anomaly. Better understanding of this transition may come from additional neutron diffraction studies to investigate how the crystal and magnetic structures evolve with temperature near $T^*$ at different magnetic fields. From the present data, it can be concluded that magnetism is involved in the transition directly or indirectly (through for example magnetoelastic coupling). The observation of an increase in resistivity upon cooling through the transition suggests that either the electronic structure is altered, or the scattering rate is increased. The former could be due to a subtle structural distortion or orbital ordering involving Ru, and the latter could be related to magnetic domain walls which form in the mixed AFM/FM state.

\begin{figure}
\begin{center}
\includegraphics[width=3.25in]{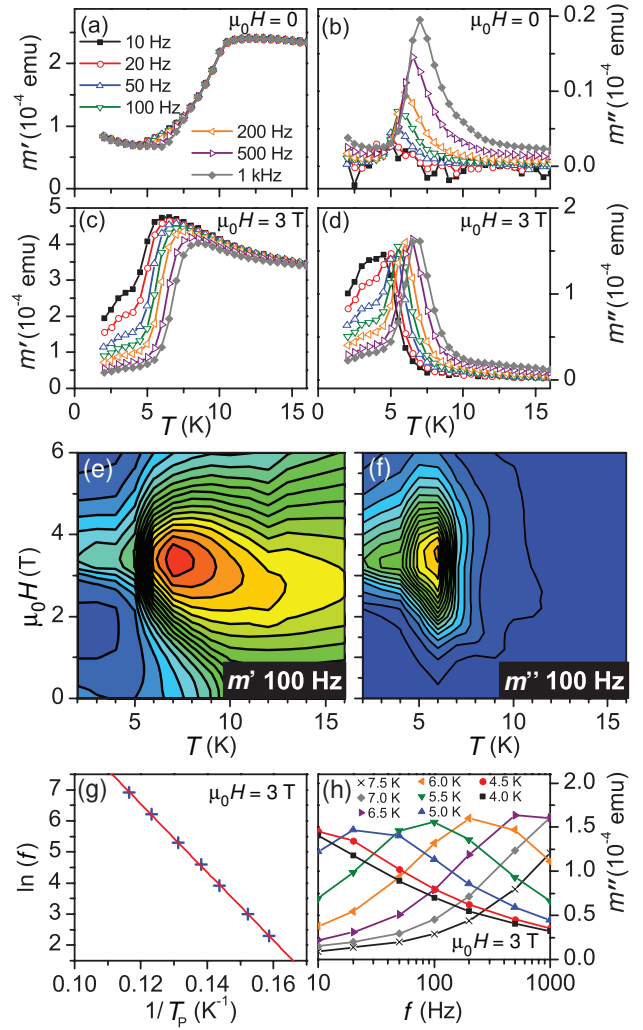}
\caption{\label{fig:acms}
(Color online) (a-d) Temperature dependence of the real ($m'$) and imaginary ($m''$) parts of the ac magnetization measured at $\mu_0H$= 0 and 3 T using frequencies indicated in panel (a). (e,f) Contour plots of the real part ($m'$) and imaginary part ($m''$) of the ac magnetization measured at 100 Hz as functions of temperature and applied dc magnetic field. (g) Arrhenius fit using the temperatures ($T_P$) at which $m'$ peaks at different frequencies for $H$ = 3 T. (h) Frequency dependence of $m''$ at $H$ = 3 T and the indicated temperatures. All measurements were conducted using a sample of mass 27.1 mg and an ac excitation field of 10 Oe.
}
\end{center}
\end{figure}

The dynamics of the low temperature magnetism in DyRuAsO were also investigated using ac magnetic susceptibility measurements. The frequency, temperature, and field dependence of the real ($m'$) and imaginary ($m''$) parts of the ac magnetization were studied near the magnetic ordering temperatures. The results are summarized in Fig. \ref{fig:acms}.

In zero applied dc field (Fig. \ref{fig:acms}a,b), the ac magnetization resembles closely the low field dc magnetization data (Fig. \ref{fig:dcmag}b), with  a sharp drop just below $T_N$. A weak frequency dependence is seen for 4.5 $< T <$ 8.0 K, well below $T_N$. A corresponding anomaly is seen in $m''$ at $\mu_0H = 0$ (Fig. \ref{fig:acms}b) and increases in both magnitude and temperature as frequency increases. The temperatures spanned by the $m''$ peak position is similar to the temperature range over which $m'$ is found to be frequency dependent.

When the applied dc field is increased to 3 T, near the meta-magnetic transition, the features noted in the zero-field data are enhanced. A strong frequency dependence is observed in $m'$ (Fig. \ref{fig:acms}c), and $m''$ is about one order of magnitude larger at 3 T (Fig. \ref{fig:acms}d) than in zero field (Fig. \ref{fig:acms}b). The temperature of the cusp in $m'$ (Fig. \ref{fig:acms}c) increases with frequency from 6.5 K at 10 Hz to 8.5 K at 1 kHz, while the peak magnitude of $m'$ decreases. The temperature at which $m''$ peaks clearly increases with frequency, while its magnitude shows a more subtle increase. Contour plots of the real and imaginary parts of the the ac susceptibility measured at 100 Hz are shown in Fig. \ref{fig:acms}e and \ref{fig:acms}f. The $m''$ data indicate that the dissipation is strongest below $T_N$ and for magnetic fields near 3 T, the region where AFM-FM competition is expected to be strongest.

The behaviors shown in Fig. \ref{fig:acms}c,d are precisely those expected for a spin-glass near its freezing temperature \cite{Spinglassbook, BinderAndYoung}. In fact, it is interesting to note the similarities of many of the behaviors of DyRuAsO at low temperatures and fields near 3 T to those of a spin glass, including frequency dependent ac susceptibility, FC-ZFC divergence in dc magnetization, and slow relaxation of the magnetic moment when the applied field is changed. However, there is no chemical disorder detected by neutron diffraction in this material.

Glass-like behavior without chemical disorder can arise from strong geometrical frustration, as realized, for example, in Dy-pyrochlore spin-ice systems \cite{Dy2Ti2O7, Dy2Sn2O7}. Similar behavior has been reported in the related Ising antiferromagnet Dy$_2$Ge$_2$O$_7$, which does not adopt the pyrochlore structure, and in which the glass-like behavior is speculated to arise from collective relaxation of short-range spin correlations \cite{Dy2Ge2O7}. The magnetic structure adopted by DyRuAsO (Fig. \ref{fig:MagStr}c) does not suggest strong frustration in this compound, due to the FM coupling within each Dy net in the \textit{ab}-plane. For purely AFM interactions, however, the structure does support geometrical frustration. Comparing the ac susceptibility and heat capacity behavior of DyRuAsO and Dy$_2$Ge$_2$O$_7$ \cite{Dy2Ge2O7}, strong similarities are observed. An important exception is the ZFC-FC divergence of dc magnetization seen in DyRuAsO (Fig. \ref{fig:dcmag}b). This is absent in Dy$_2$Ge$_2$O$_7$, and its absence is used to distinguish this material from a spin-glass. In this respect, DyRuAsO appears to behave more like a spin-glass than does the pyrogermanate. The proximity of the glass-like behavior in DyRuAsO to the metamagnetic transition suggests that domain walls separating FM and AFM domains may also play a role in the observed dynamics. Similarly, it has been suggested that AFM domain walls may contribute to the frequency dependent phenomena observed in Dy$_2$Ge$_2$O$_7$ \cite{Dy2Ge2O7}.

Further analysis of the ac magnetization data collected in a dc field of 3 T are shown in Fig. \ref{fig:acms}g,h. The relationship between the temperature at which $m'$ peaks ($T_P$) and the measurement frequency is seen to follow an Arrhenius law $f = f_0e^{-E_A/k_BT}$ (Fig. \ref{fig:acms}g), as typically seen in spin-glasses \cite{Spinglassbook, BinderAndYoung}, but not spin-ices \cite{Dy2Ti2O7-2004}. The activation energy determined from the fit is $E_A/k_B$ = 110 K, and the attempt frequency is $f_0$ = 360 MHz. A similar activation energy of 162 K is reported for Dy$_2$Ge$_2$O$_7$, which was found to correspond to the separation between the ground and first excited crystal field states \cite{Dy2Ge2O7}. The frequency dependence of $m''$ at temperatures from 4.0 to 7.5 K is shown in Fig. \ref{fig:acms}h. Plotted in this way, a peak corresponds to characteristic spin relaxation frequencies \cite{Dy2Ti2O7, Dy2Ti2O7-1}. The uniform shift in peak position with temperature is consistent with classical thermal relaxation.

\section{Summary and conclusions}

Neutron diffraction has been used to identify the magnetic ordering of Dy moments in DyRuAsO at low temperature, and how the moments respond to application of magnetic fields. The results provide a framework necessary for understanding the peculiar physical properties of this material, which is structurally and electronically related to the 1111 Fe superconductor systems. In the absence of an applied magnetic field, AFM ordering occurs at 10.5 K. The magnetic unit cell is the same as the orthorhombic crystallographic unit cell. Magnetic moments on Dy of magnitude 7.6(1) $\mu_B$ lie long the b-axis. The moments are arranged FM within sheets in the ab-plane, with AFM stacking along the c-axis. This same magnetic structure describes the diffraction data collected in a magnetic field of 2 T at $T$ = 3 K. A related FM structure was determined when the field was increased to 5 T, with moments of magnitude 7.3(3) $\mu_B$ aligned in the ab-plane. The neutron diffraction results distinguish the meta-magnetic transition occurring in DyRuAsO from the more commonly observed spin-flop.

At intermediate fields, the competition between the AFM and FM states is evident in the physical properties of DyRuAsO, and results in several unusual behaviors. Many of the field induced phenomena appear to be related to a thermal anomaly identified at $T^*$ = 3 K, which is evident in the heat capacity for magnetic fields near 3 T. The resistivity increases upon cooling through $T^*$. The dc magnetization shows a subtle inflection near $T^*$, and develops a FC-ZFC divergence slightly above $T^*$. In addition, the transport and magnetic properties develop a time dependence below $T^*$ for fields near 3 T, producing apparent thermal and magnetic hysteresis in magnetization and magnetoresistance measurements. The observed slow relaxation of magnetization, as well as frequency dependent ac magnetic susceptibility values, are reminiscent of behaviors associated with spin-glasses \cite{Spinglassbook} and Dy-based spin-ice and related materials \cite{Dy2Ti2O7, Dy2Sn2O7, Dy2Ti2O7-1, Dy2Ge2O7}. This is somewhat surprising; no chemical disorder is detected in this material, and the geometry does not indicate strong magnetic frustration \cite{McGuire-TbDyRuAsO}. Movement of magnetic domain walls related to the competing AFM and FM phases provide one possible source of dissipation.

Since the magnetism in DyRuAsO is dominated by the Dy atoms, crystalline electric field effects likely play an important role in determining the magnetic properties. Indeed, it has been noted that the energy barrier to spin-relaxation in Dy$_2$Ge$_2$O$_7$ corresponds to the energy splitting of the lowest crystal field levels \cite{Dy2Ge2O7}.The current experimental data show a saturation moment near 7 $\mu_B$ (similar to the refined ordered moment from the diffraction results), which is significantly smaller than the free ion value of 10 $\mu_B$, and the magnetic entropy determined from the heat capacity is relatively small. The Curie Weiss behavior of the magnetization for temperatures just above $T_S$ is consistent with the free ion value of the effective moment. Detailed calculations and inelastic scattering experiments (complicated by relatively strong neutron absorption by Dy) would be desirable in developing an understanding of these effects. Since this material undergoes a structural phase transition and displays meta-magnetic behavior, the dependence of the Dy crystal field levels on temperature, field, and coordination details will be required to develop a complete picture.

Iron magnetism is closely linked to the structural distortion that occurs in the isoelectronic Fe compounds \cite{Johnston-review, Lumsden-Christianson-review, McGuire-review}. In the present data, no conclusive evidence for ordered magnetic moments on Ru atoms is seen, and no magnetic order is observed between $T_N$ and $T_S$ in the absence of an applied magnetic field. The driving force for the structural transition at $T_S$, which also occurs in DyZnAsO, is identified as Dy magnetism. The distortion must alter the Dy crystal field levels, which provides some degree of magnetoelastic coupling. In addition, TbRuAsO is isostructural with DyRuAsO at room temperature and was found here to have a low temperature magnetic structure similar to DyRuAsO, but no structural distortion occurs in TbRuAsO at temperatures as low as 1.5 K.

ZrCuAsSi-type oxyarsenides incorporating heavy transition metal atoms have been relatively little studied compared to the 3$d$ metal analogues or the related ThCr$_2$Si$_2$-type arsenides. The structure type shows a large degree of chemical flexibility. Many interesting materials and behaviors have already been identified; however, many compounds and phenomena likely remain undiscovered or understudied. The surprisingly complex and glass-like magnetic behavior of DyRuAsO suggest further study of such compounds will uncover other new and interesting cooperative phenomena, and motivates further study of dynamic effects in rare-earth magnetic materials.
\\

Research sponsored by the US Department of Energy, Office of Basic Energy Sciences, Materials Sciences and Engineering Division. Neutron scattering at the High Flux Isotope Reactor was supported by the Scientific User Facilities Division, Basic Energy Sciences, US Department of Energy. The authors thank J.-Q. Yan for helpful discussions.

\bibliography{DyRuAsO}% Produces the bibliography via BibTeX.
\end{document}